\documentclass{article}

\usepackage{arxiv}
\usepackage[utf8]{inputenc}
\usepackage[T1]{fontenc}
\usepackage[hidelinks]{hyperref}
\usepackage{url}
\usepackage{booktabs}
\usepackage{amsfonts}
\usepackage{nicefrac}
\usepackage{microtype}
\usepackage{graphicx}
\usepackage[numbers]{natbib}
\usepackage{doi}

\title{SPELUNKER – Item Similarity Search Using Large Language Models and Custom K-Nearest Neighbors}

\date{August 26, 2025}

\author{ \href{https://orcid.org/0009-0006-3252-7343}{\includegraphics[scale=0.06]{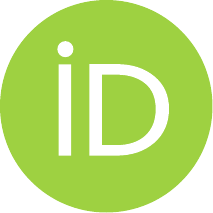}\hspace{1mm}Ana C. Rodrigues}\\
	Crab Technologies, Lda\\
	Rua do Poço, 28, 4900-519, \\
        Viana do Castelo, Portugal \\
	\texttt{ana.rodrigues@crabtechnologies.com} \\
        \And
	\href{https://orcid.org/0009-0007-3010-2672}{\includegraphics[scale=0.06]{orcid.pdf}\hspace{1mm}João Mata} \\
	Instituto Superior Técnico\\
	Lisbon University\\
        Lisbon, Portugal\\
	\texttt{joao.m.mata@tecnico.ulisboa.pt} \\
        \And
	{Rui Rego} \\
	Crab Technologies, Lda\\
	Rua do Poço, 28, 4900-519,\\
        Viana do Castelo, Portugal \\
	\texttt{rui.rego@crabtechnologies.com} \\
}

\renewcommand{\headeright}{}
\renewcommand{\undertitle}{Technical Report}
\hypersetup{
pdftitle={SPELUNKER – Item Similarity Search Using Large Language Models and Custom K-Nearest Neighbors},
pdfsubject={q-bio.NC, q-bio.QM},
pdfauthor={Ana C. Rodrigues, João Mata, Rui Rego},
pdfkeywords={LLM, KNN, vector, embeddings, similarity},
}

\begin{document}
\maketitle

\begin{abstract}
	This paper presents a hybrid system for intuitive item similarity search that combines a Large Language Model (LLM) with a custom K-Nearest Neighbors (KNN) algorithm. Unlike black-box dense vector systems, this architecture provides superior interpretability by first using an LLM to convert natural language queries into structured, attribute-based searches. This structured query then serves as input to a custom KNN algorithm with a BallTree search strategy, which uses a heterogeneous distance metric to preserve distinct data types. Our evaluation, conducted on a dataset of 500 wine reviews, demonstrates the system's effectiveness. The LLM achieved an F1-score of 0.9779 in information extraction, while also demonstrating high fidelity with a Jaro string similarity of 0.9321. When we augmented the KNN algorithm with LLM-based re-ranking, we observed a statistically significant improvement in recall (p=0.013), indicating the LLM's ability to identify and promote relevant items that align with nuanced user intent. This approach effectively bridges the gap between human language and machine-understandable item representations, offering a transparent and nuanced search capability.
\end{abstract}

% keywords can be removed

\section{Introduction}
The proliferation of digital content and e-commerce platforms has amplified the need for effective item discovery mechanisms. Traditional search methods, often relying on keyword matching or predefined taxonomies, frequently fall short when users seek items based on subjective attributes, nuanced descriptions, or abstract concepts. This limitation requires a more sophisticated approach that can understand the semantic richness of human language. 

Recent advancements in Large Language Models (LLMs) have demonstrated remarkable capabilities in natural language understanding, generation, and transformation \cite{vaswani2017attention}. Concurrently, vector space models and similarity search algorithms, such as K-Nearest Neighbors (KNN) have proven effective in identifying relationships within high-dimensional data \cite{ali2019evaluation,cover1967nearest}.  

This paper proposes a synergistic integration of these technologies to address the challenges of intuitive item similarity search. Our system allows users to express their item preferences using natural language prompts, which are then processed by an LLM to generate structured queries. These structured queries are subsequently used by a custom KNN algorithm to find similar items from a pre-indexed dataset. This methodology promises to deliver a more natural, flexible, and powerful search experience, moving beyond rigid keyword matching, to semantic understanding

\section{Methodology}
\label{sec:methods}

\begin{figure}[h!]
\centering
\includegraphics[width=16cm]{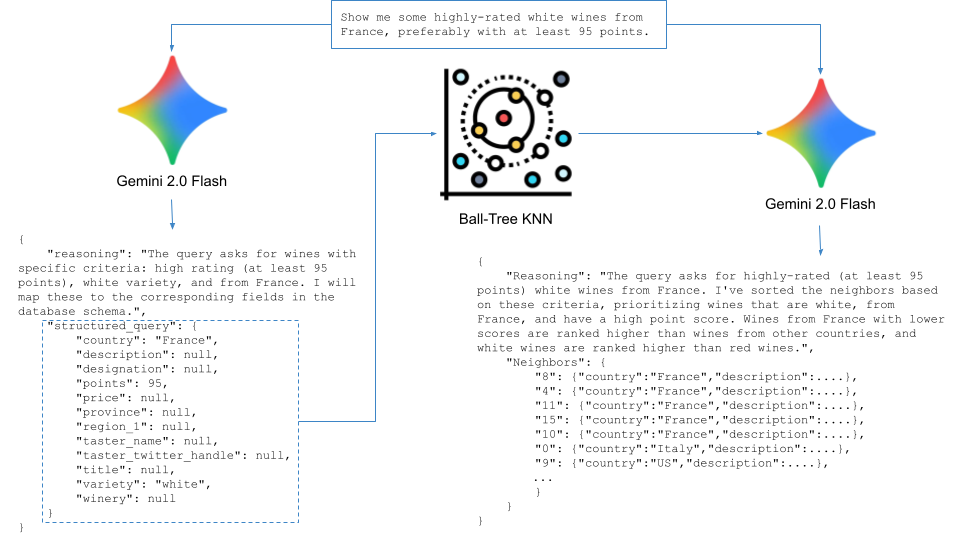}
\caption{\label{general}Diagram describing system components and information flow.}
\end{figure}

The proposed system comprises three primary components: a user interface for query input, a Large Language Model for natural language processing and structuring, and a custom K-Nearest Neighbors algorithm for similarity retrieval. The overall architecture is depicted in Figure \ref{general}.

\subsection{Dataset Pre-processing}

The dataset is preprocessed according to the data type of each variable. Numeric variables go through standard scaling. Boolean variables are converted to binary values. Categorical variables are embedded using the model all-MiniLM-L6-v2 \cite{sentenceTransformersAllMiniLML6v2}, loaded with the Sentence Transformers Python lybrary \cite{reimers2019sentence}. This model is designed to map sentences and short paragraphs into a dense vector space, making it suitable for semantic similarity tasks. 

\subsection{Custom K-Nearest Neighbors (KNN) Algorithm}

The core of the similarity search functionality is a custom K-Nearest Neighbors algorithm. A specialized distance function was developed in the following way: 

\begin{itemize}
    \item For numerical variables, Euclidean distance \cite{elmore2001euclidean} was employed to measure the straight-line distance between points in the feature space.
    \item For boolean variables, a simple match/no-match criterion was used, contributing zero distance if values are identical and 1 otherwise.
    \item For categorical variables, the distance was calculated using cosine similarity \cite{gunturclustering} on their vector embeddings.
    \item Additionally, custom weighing of different features is possible.
\end{itemize} 

To efficiently identify nearest neighbors in this multi-dimensional feature space, a custom Ball Tree search strategy \cite{dolatshah2015ball} was implemented. This data structure partitions the feature space into nested hyperspheres, allowing for rapid pruning of search areas and efficient retrieval of the 'K' most similar items. These 'K' items are then presented to the user as the search results. 

\subsection{User Interface}

The user interface is developed using Streamlit \cite{streamlit2024streamlit}, a Python library for creating interactive web applications. It provides a straightforward text box where users can input their natural language descriptions of the desired item. This simple and intuitive interface minimizes cognitive load for the user, allowing them to express their needs freely without adhering to specific keywords or query syntax.

\subsection{Large Language Model (LLM) Processing}

Upon receiving a natural language query from the user, the system forwards it to model Gemini 2.0 Flash \cite{team2023gemini}, loaded with Langchain Python library \cite{langchain2024langchain}. Gemini 2.0 Flash was chosen for its low latency and high accuracy. The LLM's primary role is to interpret the user's free-form text and extract relevant features, attributes, and semantic nuances, and subsequently to transform the extracted information into a JSON structured format.

\subsection{Performance Evaluation}

The software’s performance was evaluated in three different areas.

Retrieval performance of the KNN algorithm was measured in terms of Precision at K (Precision@K measures the proportion of relevant items among the top K items retrieved by the system) and Recall at K (Recall@K measures the proportion of all truly relevant items that are successfully retrieved by the system within the top K results). 

The accuracy of the information extraction and structuring by the LLM was evaluated in terms of the F1-score, where: 

\begin{itemize}
    \item True Positives: Attributes correctly extracted with correct values;
    \item False Positives: Attributes extracted but not in ground truth, or wrong value; 
    \item False Negatives: Attributes in ground truth but not extracted, or wrong value.
\end{itemize}

The full pipeline was evaluated in terms of query latency, which corresponds to the average time taken for the system to process a user's natural language query and return the search results. This includes the time for LLM processing (information extraction and pre-processing) and the KNN search using the Ball Tree Strategy. 

\section{Results}

\subsection{Dataset Description}

A random sample of 500 observations from the public Kaggle dataset Wine Reviews \cite{wine2017kaggle} was used for experimentation. Each observation in the dataset describes one wine review. The dataset contains 2 numerical variables (“points” and “price”) and 10 categorical variables (“country”, “description”, “designation”, “province”, “region\_1”, “taster\_name”, “taster\_twitter\_handle”, “title”, “variety”, and “winery”). 

Two testing datasets were created. To assess the accuracy of information extraction and structuring by the LLM, a set of 10 natural language queries and respective structured formats were manually constructed. To test the KNN retrieval system a random sample of 10 wine reviews was selected. For these observations, a set of 4-8 “neighbor” observations were manually selected from the remaining dataset.

\subsection{System Performance}

Information extraction from the natural language query using Gemini 2.0 Flash achieved a performance of 0.9742 Precision, 0.9826 Recall, and 0.9779 F1-score. Additionally, the correctly extracted fields reached a Jaro string similarity of 0.9321 with the ground truth.  

\begin{figure}[h!]
\centering
\includegraphics[width=12cm]{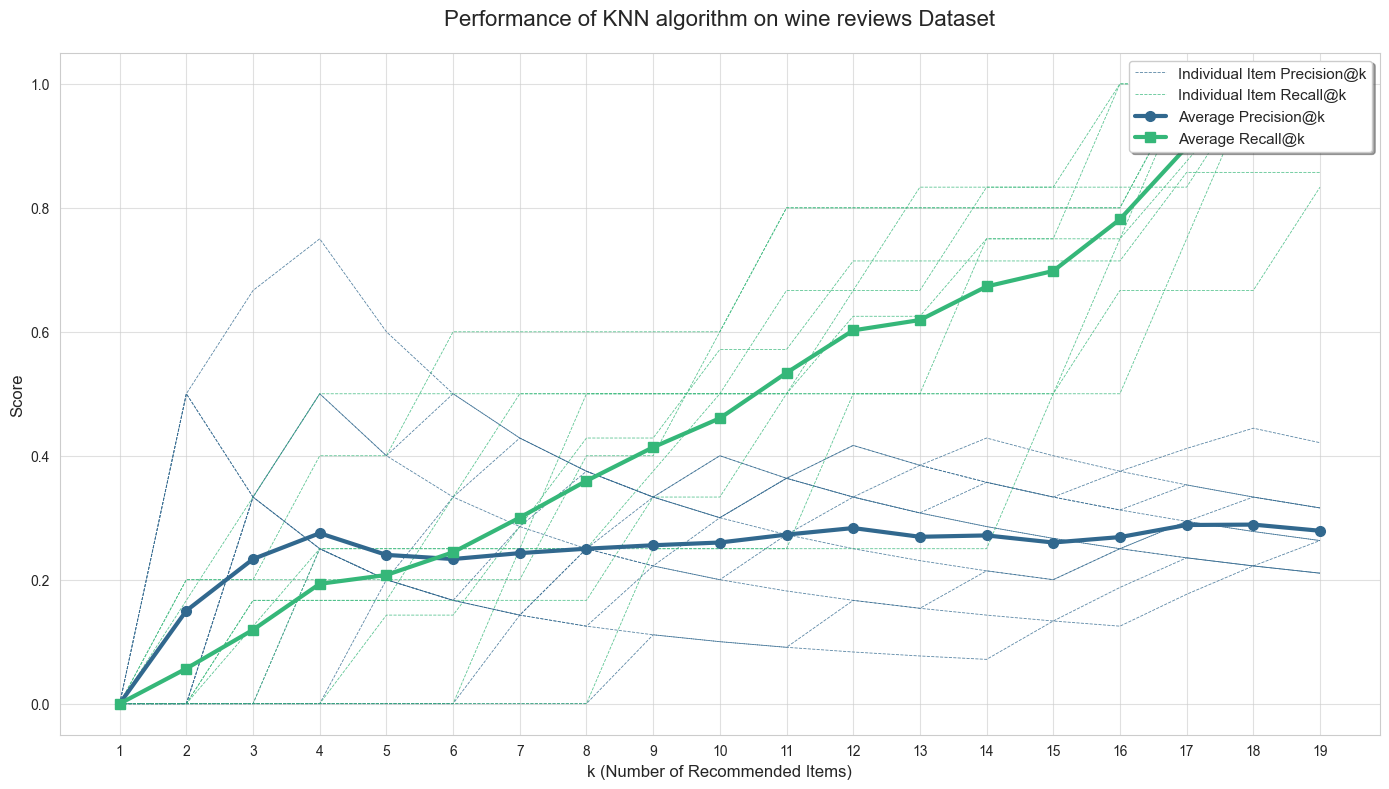}
\caption{\label{knn}KNN algorithm performance in terms of Precision and Recall at K retrieved items.}
\end{figure}

\begin{figure}[h!]
\centering
\includegraphics[width=12cm]{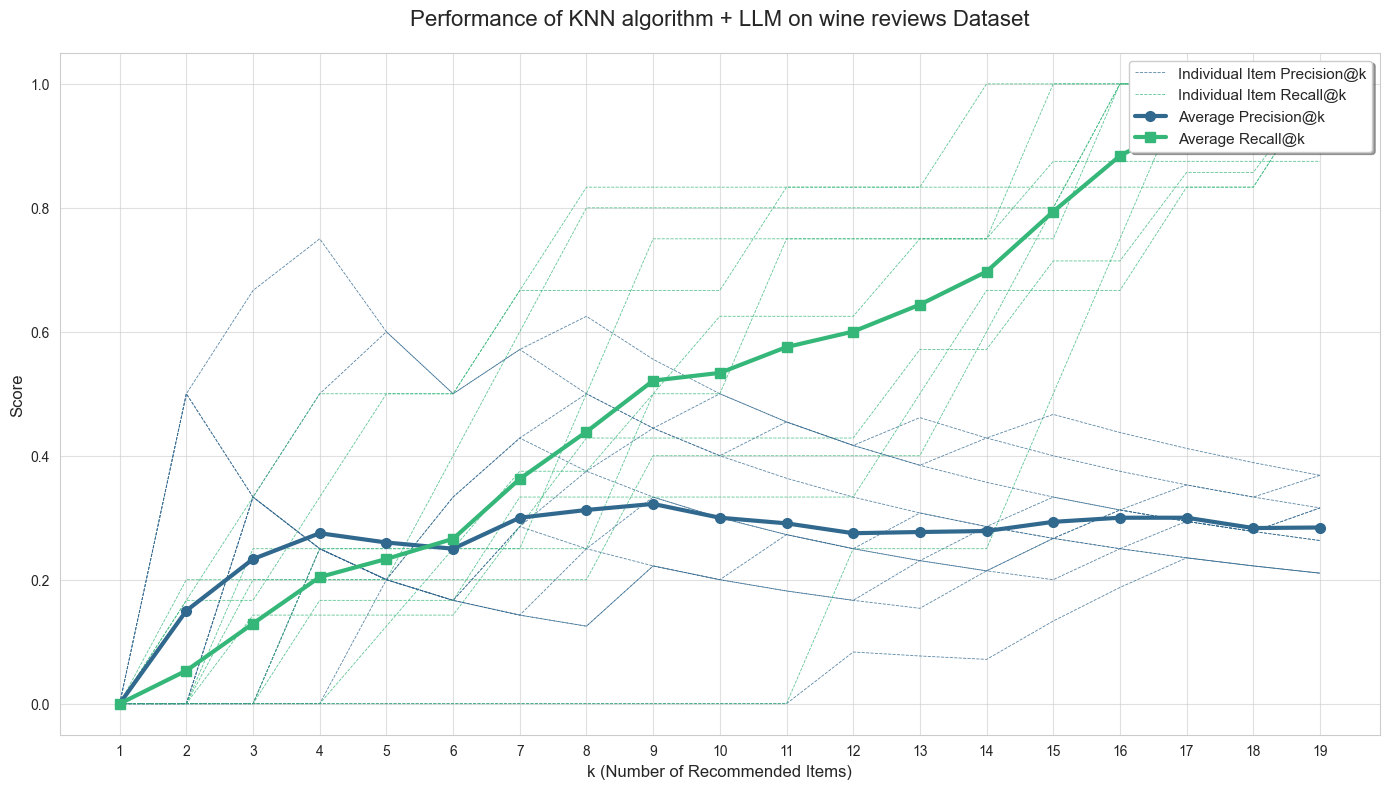}
\caption{\label{knn+llm}Performance of the KNN algorithm coupled with item sort by the LLM in terms of Precision and Recall at K retrieved items.}
\end{figure}

\begin{figure}[h!]
\centering
\includegraphics[width=12cm]{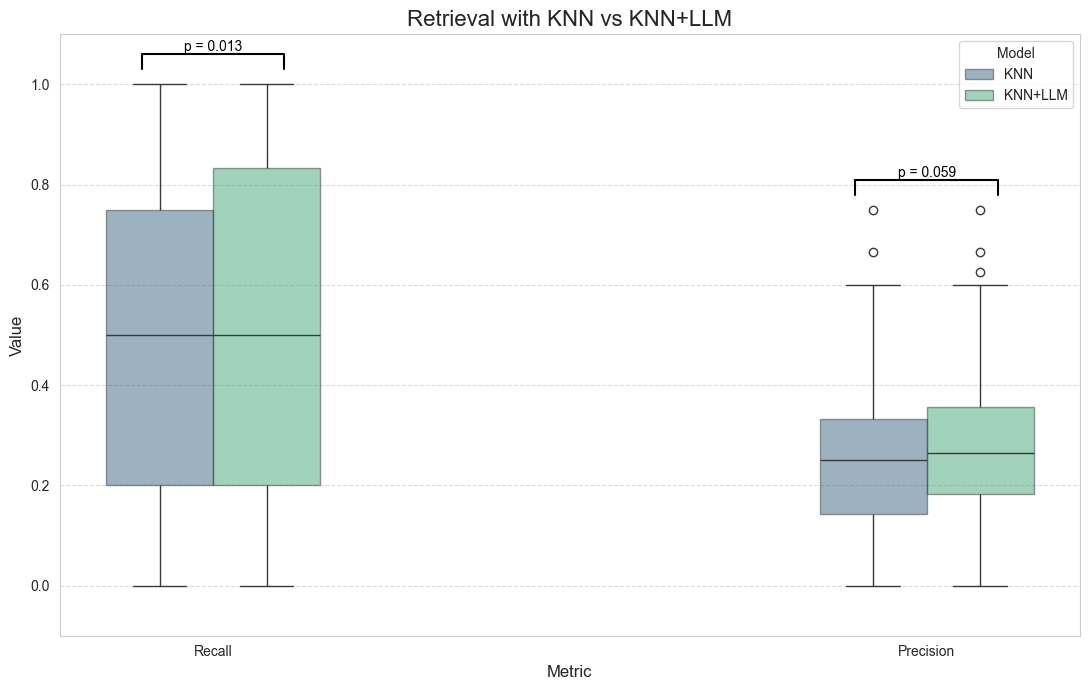}
\caption{\label{knnvsllm}Distribution of Precision and Recall performance, colored by methodology: KNN algorithm alone in blue, and KNN algorithm coupled with item sort by the LLM (KNN+LLM) in green.}
\end{figure}

The performance of the KNN algorithm alone and together with the LLM is depicted in Figures \ref{knn} and \ref{knn+llm}, respectively. The latter strategy (KNN + LLM) shows a slightly higher performance, though the difference is only statistically significant for recall (Figure \ref{knnvsllm}). 

Overall system latency was on average 18.24 seconds.

\section{Discussion}

The Spelunker system represents a successful proof-of-concept for a hybrid information retrieval architecture that synergistically combines neuro-linguistic interpretation with symbolic search. The core achievement of this work is the effective bridging of the gap between unstructured, high-level human intent and the structured, precise queries required by a database. This is accomplished through a two-stage process where an LLM acts as both a pre-retrieval query structurer and a post-retrieval results refiner. 

The Gemini 2.0 Flash model demonstrated exceptional performance in translating free-form natural language into structured JSON. The reported F1-score of 0.9779 indicates that the model is highly proficient at identifying the correct attributes (keys) and their corresponding values from user prompts. Whereas the high Jaro string similarity of 0.9321 demonstrates that it does so with high orthographic fidelity, ensuring that the structured query is not just semantically aligned but also syntactically compatible with the target database. 

A statistically significant improvement in recall (p=0.013) when augmenting the KNN search with LLM re-ranking was observed, while precision remained statistically unchanged (p=0.059). The LLM's re-ranking, guided by its internal reasoning about the holistic intent of the query, can capture priorities and semantic relationships not explicitly encoded in the distance metric. The observed increase in recall signifies that the LLM is identifying items that the KNN ranked lower and promoting them because they strongly align with the user's implicit intent. The stability of precision suggests that the initial candidate pool from the KNN is already of high quality; the LLM's primary contribution is not filtering out irrelevant items but rather re-ordering and enriching the existing set to better reflect the user's nuanced request. 

The architecture of Spelunker offers a compelling alternative to the currently dominant paradigm of end-to-end dense vector retrieval [13]. In typical dense retrieval systems, a user's query is embedded into a single high-dimensional vector and compared against a vector index of items, often using cosine similarity. Alternatively, Spelunker's design is able to address some drawbacks of the dense vector retrieval approach. 

First, Spelunker provides superior interpretability and controllability, since the system can provide an explicit explanation for the match (e.g., "This wine was recommended because its country is 'France' and its point score is '95'"). This stands in stark contrast to the "black box" nature of pure vector similarity. 

Second, the system's architecture is adept at handling heterogeneous data. The custom KNN algorithm's use of distinct, appropriate distance metrics for different data types—Euclidean for continuous numerical values, overlap for booleans, and cosine similarity for high-dimensional text embeddings—is a key strength. This approach preserves the inherent nature of the data. Forcing all item attributes including precise numerical values such as price or points, as well as discrete categories into a single dense vector can lead to significant information loss.  

Despite its promising results, the current implementation of Spelunker has several limitations. Firstly, the high query latency (18.24 seconds) is discouraging for a real-time, interactive user application. Secondly, there are concerns about scalability. As the system's performance was evaluated on a small random sample of 500 items, its behavior on datasets containing millions of items, which is common in e-commerce and content platforms, is unknown. Thirdly, the generalizability of the evaluation is constrained by the small, manually curated test sets. While the use of 10 handcrafted queries and manually selected neighbors provides a strong initial signal of the system's capabilities, it lacks the statistical power to make broad claims about its performance across a diverse range of queries and user intents. Finally, the system in its current form is static. It operates on a fixed database schema and uses static weights for different features in the KNN distance calculation. It cannot dynamically adapt to new data types or, more importantly, adjust the relative importance of different features based on the specific context of a user's query. 

Future work will include system latency optimization, by comparing different Approximate Nearest Neighbors Algorithms such as Graph-based Hierarchical Navigable Small World algorithm (HNSW) \cite{malkov2018efficient}, ANNOY \cite{annoy}, and KDTree \cite{friedman1977algorithm}. Furthermore, the system’s scalability will be tested on larger, more diverse datasets. Finally, we will explore a dynamic weighting mechanism that can adjust the relative importance of different features based on the specific context of a user's query.

\section{Conclusion}

The Spelunker system successfully demonstrates a novel architecture for item similarity search that harmonizes the natural language understanding of LLMs with the structured efficiency of a custom KNN algorithm. The high accuracy of the LLM in transforming unstructured queries into structured, database-compatible formats, combined with the statistically significant improvement in retrieval recall from LLM-based re-ranking, validates the core hypothesis of this work. Our approach offers enhanced interpretability and superior handling of heterogeneous data compared to pure vector-based retrieval methods. 

\section{Acknowledgements}

This work received funding from Crab Technologies, Lda \cite{crab}.

\bibliographystyle{unsrtnat}
\bibliography{template}

\end{document}